\documentclass[a4paper,11pt]{article}
\pdfoutput=1 

\usepackage{jinstpub} 

\usepackage{lineno}

\title{\boldmath Initial performance assessment and calibration techniques of the production monolithic silicon pixel detector ASIC for the detection of electromagnetic showers of TeV energy in the FASER experiment at the LHC}

\author[a,b,1]{Rafaella Eleni Kotitsa,\note{Corresponding author.}}

\affiliation[a]{CERN,\\Espl. des Particules 1/1211, 23 Genève, Switzerland}
\affiliation[b]{Departement de Physique Nucleaire et Corpuscolaire (DPNC), University of Geneva,\\Quai Ernest-Ansermet 24, 1205 Genève, Switzerland}

\emailAdd{rafaella.eleni.kotitsa@cern.ch}

\abstract{The FASER experiment at the LHC aims to detect new, long-lived fundamental particles. A tungsten-silicon (W-Si) preshower detector is being developed to enhance the experiment's ability to distinguish between closely spaced photon pairs with energies on the order of TeV and separations as small as 200 \textmu m. This detector will use a monolithic silicon pixel sensor fabricated with 130 nm SiGe BiCMOS technology. Initial tests were conducted on the production ASIC, after evaluating the performance of the preproduction one. The first tests on the production ASIC indicate that the essential components of the monolithic silicon pixel detector are operating as expected. Two calibration methods were developed to reconstruct the charge of the FASER preshower detector, from the ASICs' ADC values.}


\arxivnumber{} 

\collaboration[c]{on behalf of FASER collaboration}

\proceeding{25$^{\text{th}}$ International Workshop on Radiation Imaging Detectors (iWoRiD)\\
  30 June - 4 July 2024\\
  Lisbon, Portugal}

\begin{document}
\maketitle
\flushbottom

\section{The FASER experiment and the preshower upgrade}
\label{sec:FASER}
\begin{figure}[htbp]
\centering 
\includegraphics[width=1\textwidth]{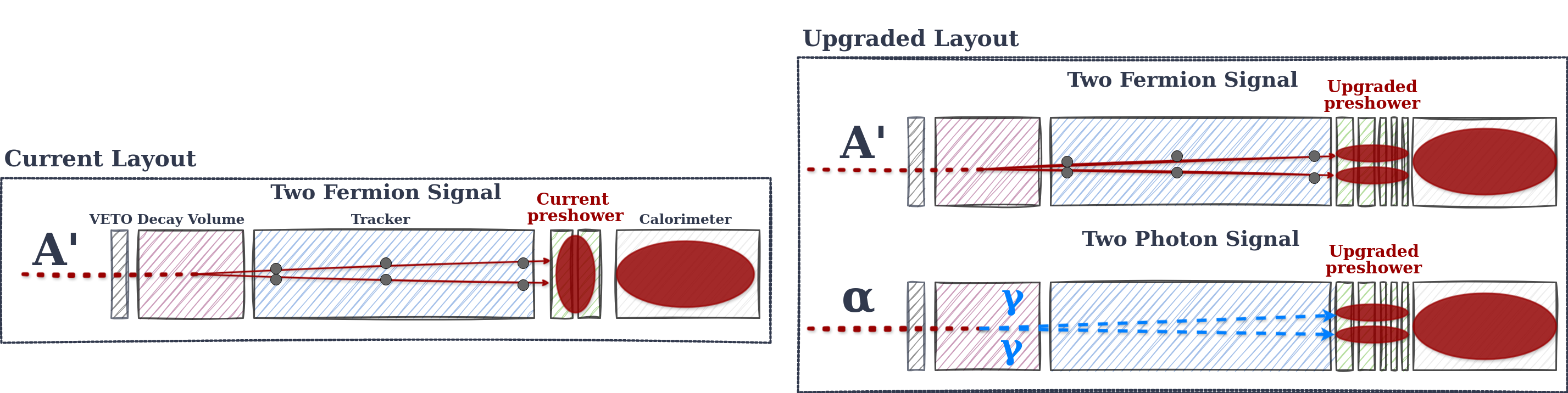}
\caption{\label{fig:FASER} Left side: The current layout of FASER. This illustration shows the detector signature of a dark photon ($A'$) decaying into an electron-positron pair within FASER's decay volume, with the $A'$ entering the detector from the left. Right side: The upgraded layout of FASER, shows the same signal for the current layout of the detector, but also the new preshower detector signature of a Long-Lived Particle ($\alpha$) decaying into a photon-photon pair inside FASER's decay volume, with the $\alpha$ also entering the detector from the left ~\cite{preshower}.}
\end{figure}

\begin{figure}[htbp]
\centering 
\includegraphics[width=1\textwidth]{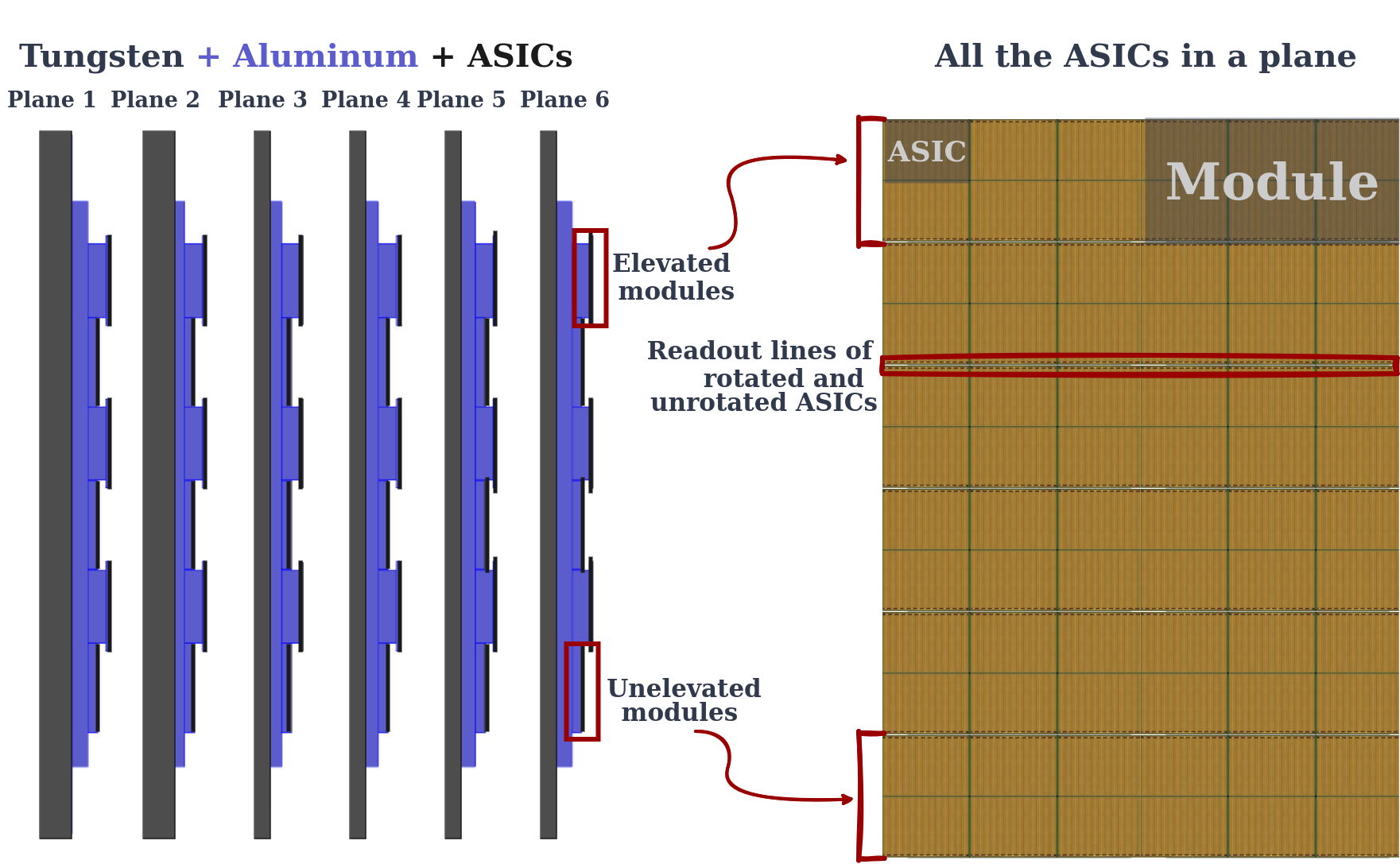}
\caption{\label{fig:preshower_G4} The left side presents the side view of the upgraded preshower detector in the Allpix Squared~\cite{allpix} simulation ~\cite{allpix}. The Geant4 visualization uses distinct colors to represent the tungsten, aluminum cooling plate, and aluminum module base. The ASICs on top of the modules are shown in black, along with the rows of elevated modules. The right side features a sketch illustrating the number of ASICs per plane, with each module containing 6 ASICs, totaling 72 ASICs per plane. The figure also shows the rotation of the ASICs within each module, optimized to maximize the active area. These design details are fully implemented in the Allpix Squared~\cite{allpix} simulation on the right. The ASIC for making this illustration is from Cadence~\cite{cadence} and taken by ~\cite{lor}.}
\end{figure}

The Forward Search Experiment (FASER) at the Large Hadron Collider (LHC), aims to detect light and extremely weakly interacting particles generated at zero angle by proton-proton collisions. These particles are hypothesized by various models beyond the Standard Model, which seek to address some unresolved questions in physics, including the nature of dark matter, the origin of neutrino masses, and the observed matter-antimatter asymmetry in the universe~\cite{faser_web}. If such particles exist, they might be generated by proton-proton collisions at the ATLAS detector’s interaction point within the LHC. Capable of traveling significant distances through rock and concrete without interacting, these particles are expected to decay into detectable particles within FASER's decay volume, as depicted in Figure \ref{fig:FASER}. FASER is situated 480 meters from the ATLAS experiment's proton collision point at the LHC, aligned with the beam collision axis, and it has been taking data during LHC Run 3. An upgrade of the FASER detector is foreseen at the end of 2024. 
The current FASER preshower, composed of 2 radiation lengths of tungsten absorber and scintillating detectors, lacks sufficient granularity to discriminate particles on the beam's transverse plane. To overcome this limitation, a new high-resolution tungsten-silicon preshower will be installed in front of the calorimeter to enhance the existing setup~\cite{preshower}. This enhanced detector will enable the differentiation of closely spaced photons with energies ranging from 100 GeV to several TeV, such as those associated with ALP signatures (Figure \ref{fig:FASER}), from the neutral particle background. 
The upgraded preshower configuration will consist of six layers of tungsten absorber, with planes of monolithic silicon pixel detectors in between, as illustrated in Figure \ref{fig:preshower_G4}. The tungsten layers, totaling 6 radiation lengths in thickness, will facilitate efficient photon conversion and electromagnetic shower development, allowing the preshower detector to effectively sample and reconstruct two electromagnetic showers and accurately determine their cores. The first two tungsten layers will be thicker to allow early photon conversion.
The detector planes will consist of 12 modules, distributed across 2 columns 6 rows per plane. The modules will be placed on cooling plates, with a 2 mm overlap between adjacent modules to reduce inactive areas. Each module will house six monolithic silicon pixel detector ASICs, organized in a 2$\times$3 grid (Figure \ref{fig:preshower_G4}). The preshower installation is planned to start in December 2024, with data acquisition extending through the end of LHC Run 3~\cite{preshower} and continuing during the High Luminosity LHC program (HL-LHC)~\cite{preshower}.

\section{The FASER ASIC}
\label{sec:preshower}
The silicon monolithic ASIC for the new FASER preshower detector was developed through a collaboration between University of Geneva, CERN, and Karlsruhe Institute for Technology, and is manufactured using 130 nm SiGe BiCMOS technology by IHP Microelectronics on high-resistivity wafers~\cite{preshower}. Figure \ref{fig:asics_cadence} presents the layouts of the preproduction ASIC (left), received in 2022, and the production ASIC (right), received in 2024. The production ASIC, measuring $2.2\times1.5$ cm$^2$, consists of  $208\times128$ hexagonal pixels, with 65 µm side ~\cite{preshower}. The pixel matrix is organized into 13 individual readout blocks known as supercolumns. The front-end electronics are integrated within the pixels and it uses SiGe Heterojunction Bipolar Transistors, aiming for a time resolution of 300 ps ~\cite{preshower}. The power consumption is maintained below 150 mW/cm$^2$ ~\cite{preshower}.
\begin{figure}[htbp]
\centering 
\includegraphics[width=1\textwidth]{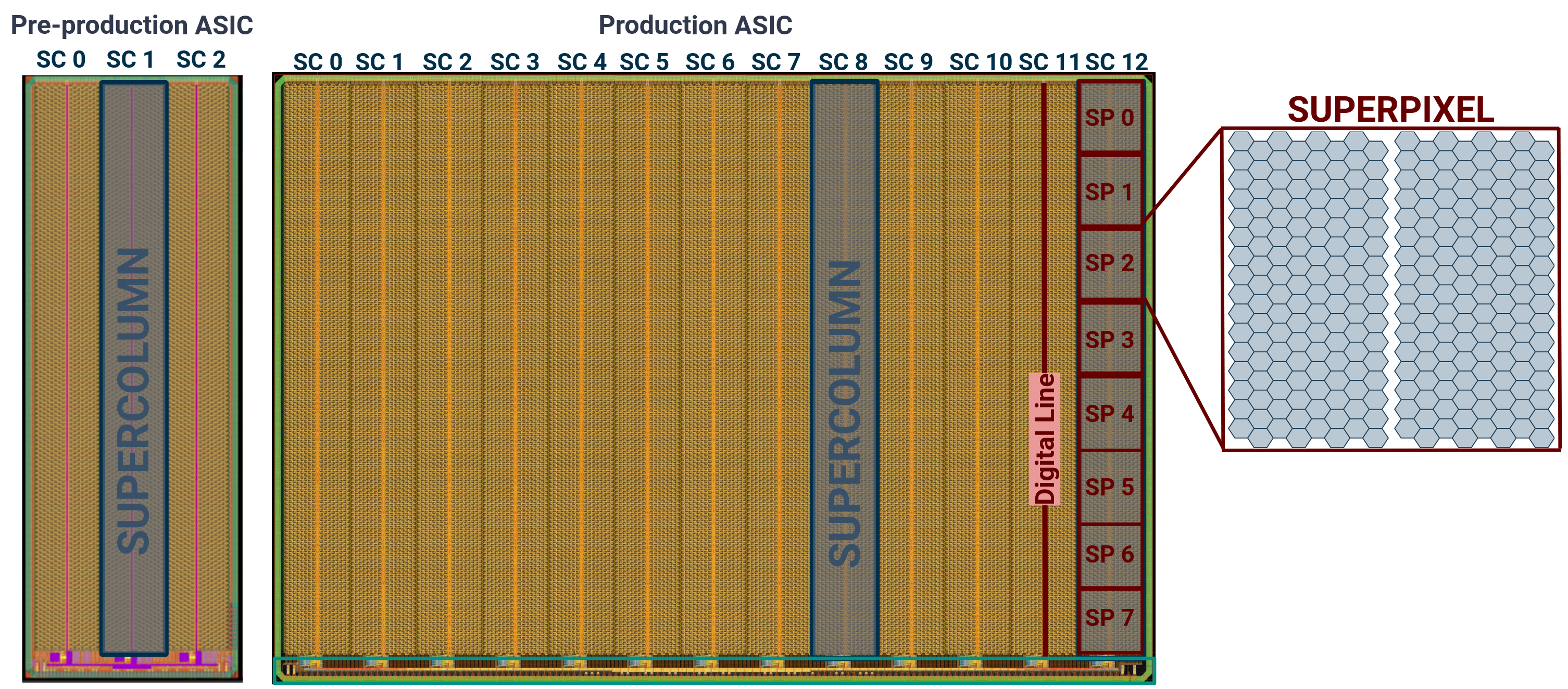}
\caption{\label{fig:asics_cadence} Left side: The preproduction prototype ASIC. It consists of three supercolumns and 8 superpixels. Each supercolumn has a thick digital line in the middle. Right side: The production ASIC, almost 4 times bigger than the preproduction one, with 13 supercolumns and 8 superpixels. Eaxh superpixel consists of 256 pixels. The ASICs for making this illustration is from Cadence~\cite{cadence} and taken by~\cite{lor}.}
\end{figure}
Simulations conducted with Allpix Squared~\cite{allpix} for the upgraded preshower detector indicate that 97\% of the events generated by TeV-scale photons have a maximum pixel charge in the final detector plane of less than 64 fC ~\cite{raf}. The front-end electronics can measure charges from 0.5 fC to 64 fC, which are digitized using 4-bit flash ADCs distributed within the supercolumn readout block. Each supercolumn is subdivided into 8 superpixels, each equipped with its own flash ADC~\cite{preshower}.
\begin{figure}[htbp]
\centering 
\includegraphics[width=1\textwidth]{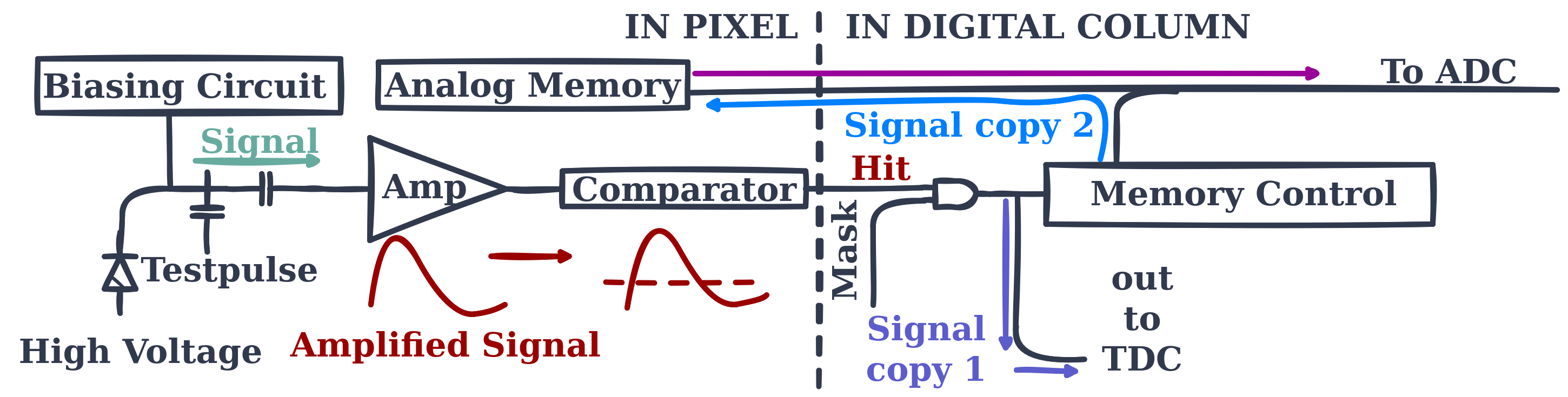}
\caption{\label{fig:asics_cadence} Diagram of a pixel layout and the process after a signal is generated \cite{lor}. Each row contains 16 pixels, with preamplifiers, discriminators, bias circuitry, and analog memory. The preamplifier generates a time-over-threshold (ToT) signal proportional to the logarithm of the input charge. This diagram is remade and adapted from ~\cite{lor}.}
\end{figure}
When a particle hits the sensor, the charge information is stored in the pixel's analog memory, while fast-OR signals are sent to the digital periphery to trigger the ASIC readout. The Time-to-Digital Converters (TDC) digitize the time of arrival of the fast-OR signals. Upon receiving the trigger, the digital periphery reads out the chip column by column. During readout, the stored charge is digitized using a 256-to-1 analog multiplexer and the 4-bit flash ADC, and then transferred to the periphery~\cite{preshower}.

\section{Test of the production ASIC}
\begin{figure}[htbp]
\centering 
\includegraphics[width=1\textwidth]{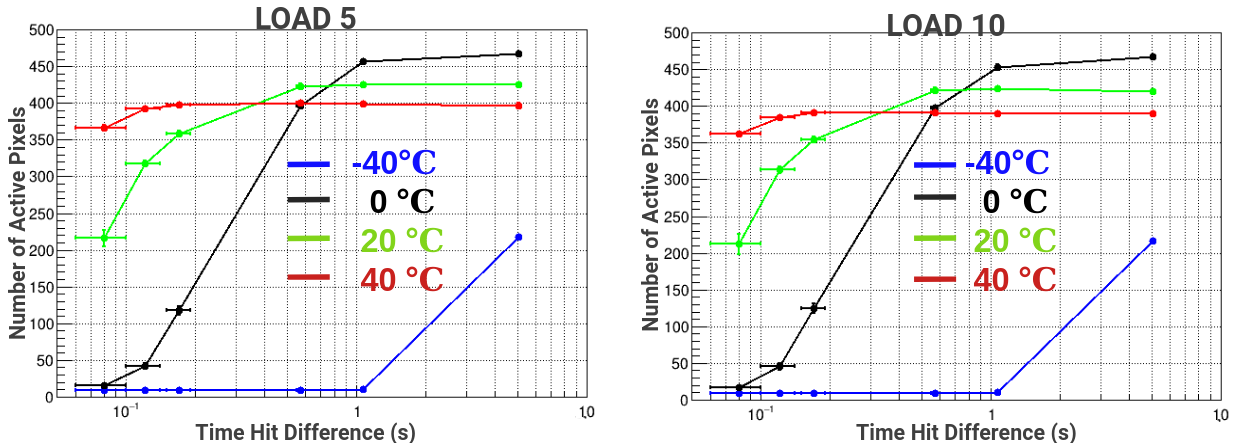}
\caption{\label{fig:climate_chamber} Climate chamber measurements from the preproduction ASIC show two working points: 0.3 µA (left) and 0.6 µA (right). Different colors represent temperatures of -40°C, 0°C, 20°C, and 40°C. The x-axis shows the time hit difference, indicating waiting times for data acquisition (10 ms, 50 ms, 100 ms, 1 s, and 5 s), while the y-axis displays the number of pixels indicating leakage.}
\end{figure}

\begin{figure}[htbp]
\centering 
\includegraphics[width=1\textwidth]{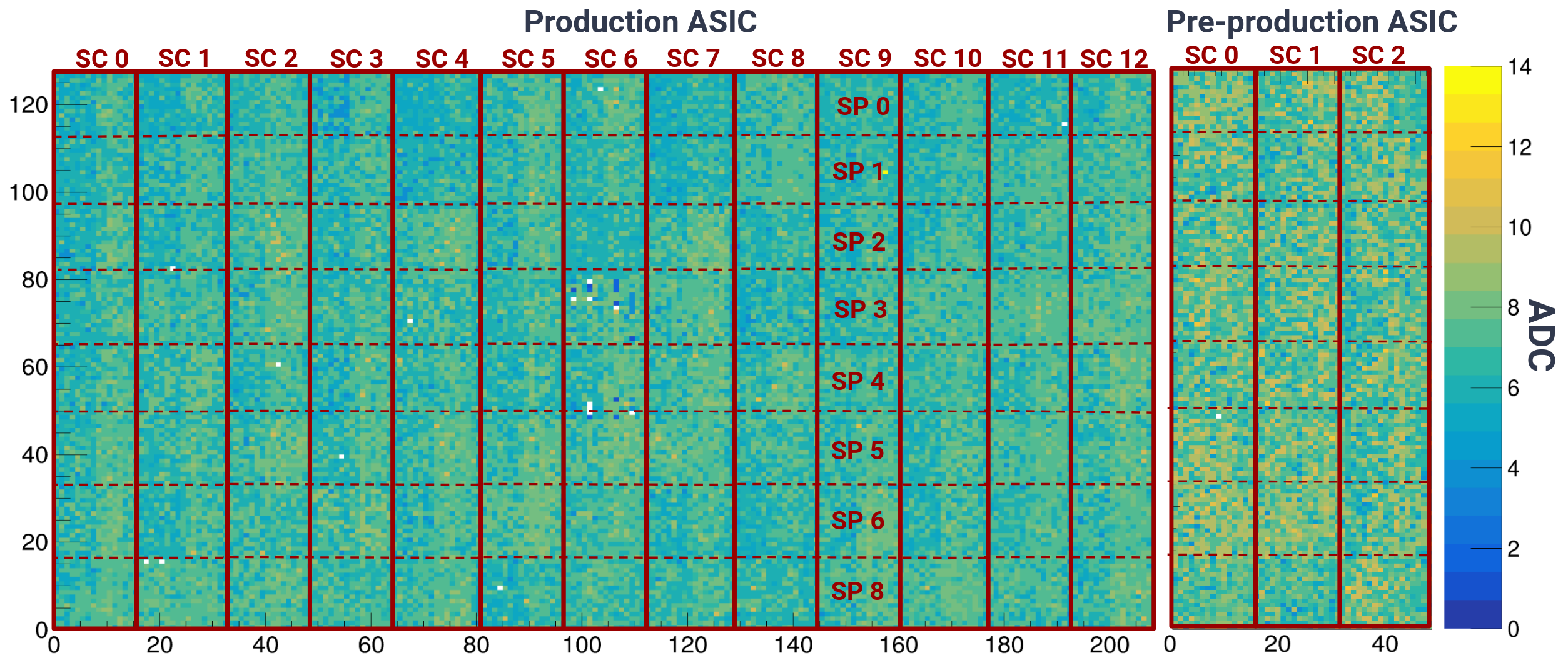}
\caption{\label{fig:adc_missmatch} The ADC maps of the production ASIC (left) and the preproduction ASIC (right) are displayed after the injection of a uniform 14 fC charge. The figure also illustrates the division of supercolumns and superpixels for both ASICs.}
\end{figure}

\begin{figure}[htbp]
\centering 
\includegraphics[width=1\textwidth]{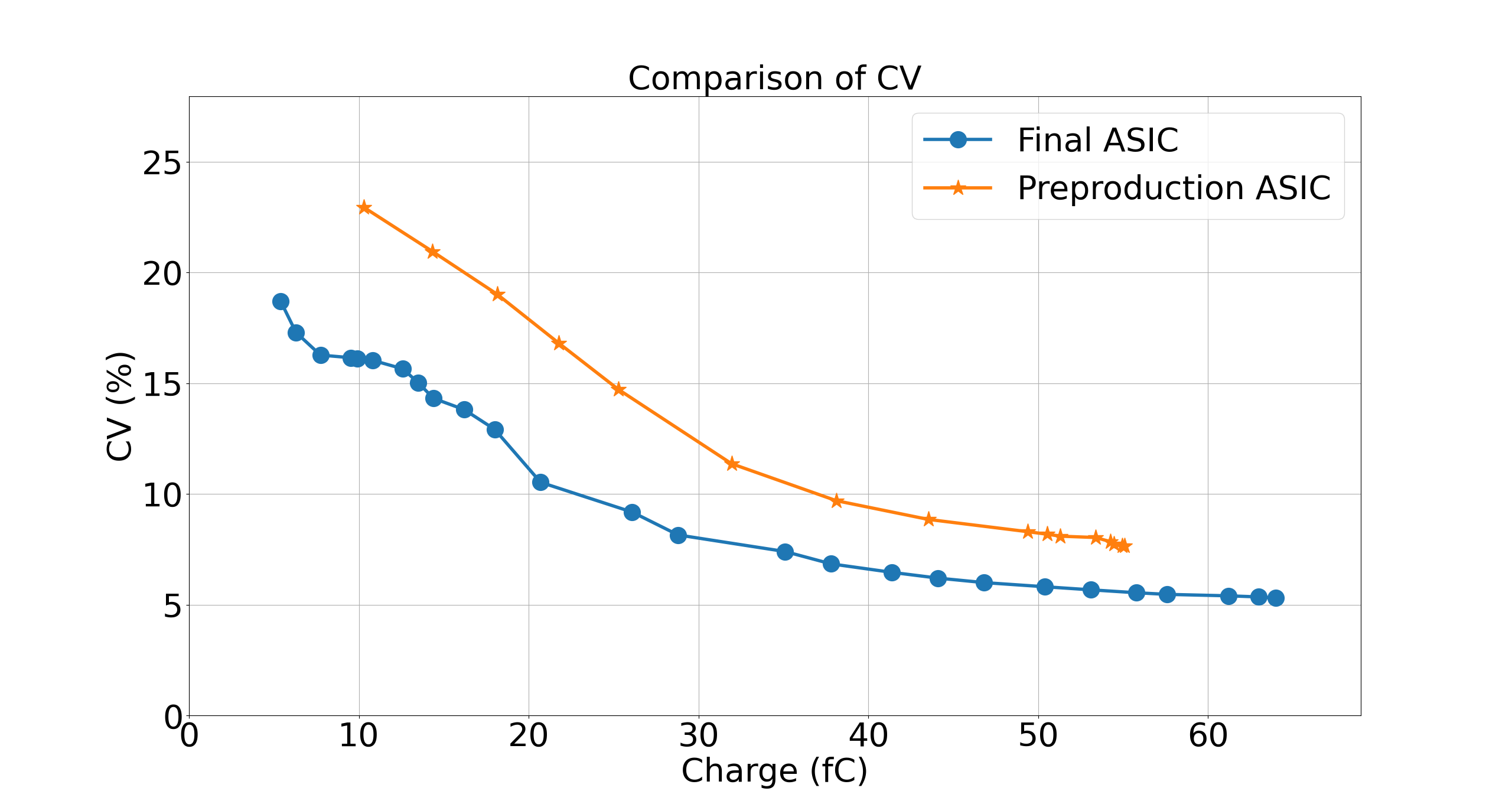}
\caption{\label{fig:coeff_variation_adc} The comparison of coefficient of variation (CV), $CV = (\sigma / \mu) \times 100 \%$, for the ADC values between the preproduction ASIC, in orange, and the production ASIC, in blue, for the same charge values shows an improvement of approximately 23\%.}
\end{figure}

Before the final production run, functionality tests of the analog memory, flash ADC, and analog multiplexer were conducted using the preproduction ASIC. Initial tests revealed some pixels remained constantly active, likely due to leakage current in the analog memories, necessitating further measurements in a climate chamber. In the test, all pixels except one were masked, and a testpulse was sent, forcing the readout in the whole ASIC. Measurements were conducted at -40 Celsius, 0 Celsius, 20 Celsius, and 40 Celsius with varying waiting times to assess leakage over time at different temperatures. The results, shown in Figure \ref{fig:climate_chamber}, confirmed leakage and demonstrated that the pedestal is temperature-dependent.
As shown in Figure \ref{fig:climate_chamber}, at -40 Celsius, the hit rate dropped significantly after 1 second. With a capacitance of 1.9 pF and bin voltage of 0.030 V, the leakage current was calculated to be 60 fA. While this delay may not impact high-rate experiments, it could affect FASER's low-rate performance. The leakage behavior was consistent across different amplifier working points, indicating it is inherent to the ASIC rather than dependent on amplifier conditions.
\label{sec:test}
The leakage was traced from the transistor gates towards the analog multiplexer, prompting The issue arose from analog memories leaking stored charge to ground via the transistor gates, causing the memory to drift instead of holding its charge~\cite{lor}. The leakage compensation added was a transistor, acting as a switch. Connects the memory node to Vdd during the "waiting for readout" phase, ensures any leakage current flows towards Vdd and prevents the memories to leak towards the ground~\cite{lor}. Later, during the readout phase, the switch opens, allowing the memory to be read by the flash ADC. In the final ASIC, no leakage was detected.
To evaluate the functionality of the ADC, varying charge values were injected into the ASIC matrix, first in the preproduction ASIC and later in the production ASIC once it became available. Figure \ref{fig:adc_missmatch} shows the final ASIC with 13 supercolumns after injecting 14 fC into all pixels of the matrix, using a specific rolling mask pattern. In the preproduction ASIC, ADC mismatch was observed across the pixels, with ADC values varying by more than 1 LSB from pixel to pixel despite identical charge injections. This variation in ADC values could undermine the accuracy of charge measurements, which in turn may impact the overall performance of the preshower detector.
No correlation was detected between the ADC values and the pixel positions. Hence the variation is primarily due to mismatches in the sizes of the feedback transistors within the front-end amplifier~\cite{lor}. Figure \ref{fig:coeff_variation_adc} compares the coefficient of variation (CV), defined as $CV = (\sigma / \mu) \times 100 \%$, for each charge value between the preproduction ASIC (shown in orange) and the production ASIC (shown in blue).
More specifically, the CV values shown in Figure \ref{fig:coeff_variation_adc} were calculated for all corresponding maps, such as the one in Figure \ref{fig:adc_missmatch}. The standard deviation and average for all pixels were then determined to quantify the variability between them for the same charge value. Figure \ref{fig:coeff_variation_adc} clearly demonstrates that the mismatch in the production ASIC is significantly improved and reduced by approximately 23\%, which is attributed to increasing the size of the preamplifier transistors from 1 µm$^2$ to 4.5 µm$^2$.

\section{Calibration}
\label{sec:calibration}
\begin{figure}[htbp]
\centering 
\includegraphics[width=1\textwidth]{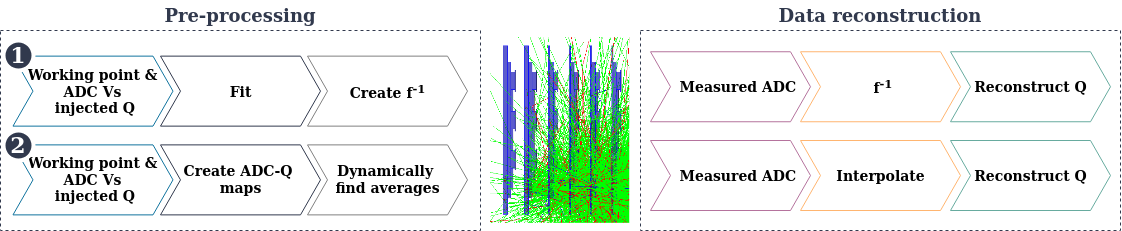}
\caption{\label{fig:flowchart} Flowchart illustrating the two charge calibration methods for the preshower detector. The top row (1) represents the "fitting" method, where the working point and ADC versus injected charge (Q) relationship are established, followed by fitting the data to create an inverse function $f^{-1}$. This function is then used in the data reconstruction phase to convert measured ADC values into reconstructed charge values. The bottom row (2) depicts the "averaging" method, which involves creating ADC-Q maps and dynamically finding averages. In the data reconstruction phase, measured ADC values are interpolated to obtain the reconstructed charge. Both methods start with the same preprocessing steps but differ in their approach to data reconstruction. The simulation in the beginning of the data reconstruction part, is a Geant4 visualization from simulations of the preshower detector from Allpix Squared~\cite{allpix}.}
\end{figure}

\begin{figure}[htbp]
\centering 
\includegraphics[width=1\textwidth]{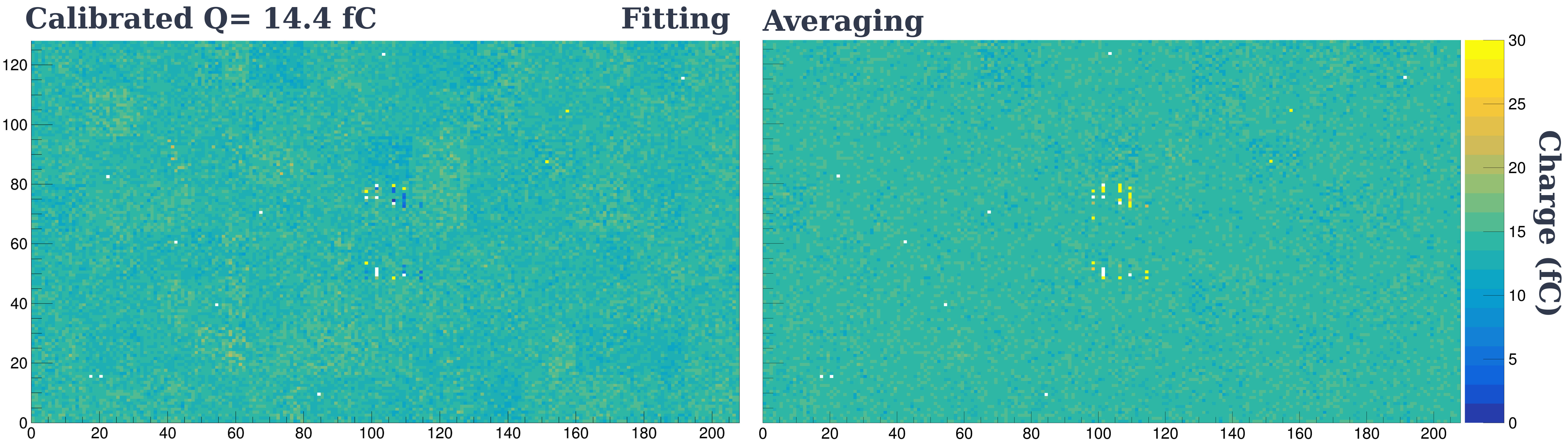}
\caption{\label{fig:calibration} Reconstructed ADC maps of the final ASIC after injecting 14.4 fC into all pixels. The left side shows the results from the "fitting" method, with an average reconstructed charge value of 13.9 fC. The right side displays the results from the "averaging" method, which yields a closer average reconstructed charge value of 14.17 fC.}
\end{figure}

The calibration of the preshower detector involves calibrating each of the 432 ASICs in the preshower detector through four main steps and is designed to optimize the performance of each ASIC, ensuring accurate detection and processing of signals. This process begins with setting a global threshold, per ASIC, during the threshold scan, where various thresholds are tested to minimize noise while maintaining sensitivity to real signals. Noisy pixels are identified and masked to prevent interference with data collection. A subsequent noise scan confirms the threshold's effectiveness by running the ASIC at the selected level, further masking any additional noisy pixels. The final step is the charge calibration, where the response of each pixel to different charge injections, through the testpulse circuit, acquired. Using a rolling mask pattern, different charges are injected into selected pixels, and the resulting ADC values are recorded. In the Figure \ref{fig:flowchart}, the flowchart of the calibration methods is depicted. 
Two potential methods are under consideration for charge calibration of the preshower detector. The first method, referred to as the "fitting" method, is depicted at the top of Figure \ref{fig:flowchart}, while the second method, the "averaging" method, is illustrated in the bottom part of the flowchart. Both methods involve the calibration of individual pixels, meaning each pixel is calibrated based on its own behavior. Up to this point, both methods follow the same initial three steps, including the acquisition of calibration data, which are the Q-ADC values mentioned earlier.
These Q-ADC values are used to create a "calibration curve" for each pixel, mapping the relationship between the injected charge and the output ADC value. As shown in the beggining of the flowchart of Figure \ref{fig:flowchart}, this process is identical for both methods. In the fitting method, the next step involves fitting the Q-ADC data to generate a function $Q=f(ADC)$ that describes the individual behavior of each pixel. This function is then dynamically adjusted to find its limits, and upon inversion, it is ready to be applied during the experiment. When actual data are collected, the measured ADC value is input into the inverted function to obtain the reconstructed charge. An example of the reconstructed charge from the fitting method, based on the ADC map of the final ASIC with a charge injection of 14.4 fC (as shown in Figure \ref{fig:adc_missmatch}), is presented on the left side of Figure  \ref{fig:calibration}.
The second method, known as the "averaging" method, involves creating Q-ADC maps, followed by the generation of a lookup table. This method dynamically identifies points where the ADC shows the same value for the same charge injection (i.e. saturation points). Upon receiving the measured charge, a sophisticated interpolation is performed, or the charge is assigned to the corresponding averaged value, resulting in the reconstructed charge. The outcome of this calibration method, applied to the same ADC map shown in Figure \ref{fig:adc_missmatch}, is displayed on the right side of Figure \ref{fig:calibration}. The average reconstructed charge value of 14.4 fC in Figure \ref{fig:calibration} is 13.9 fC for the fitting method, while the averaging method yields 14.17 fC, which is closer to the actual injected charge. Both methods perform relatively well in reconstructing various charges within the 0.5 to 64 fC range. However, the comparison between the fitting and averaging methods for charge calibration of the preshower detector reveals that the fitting method becomes more sensitive near saturation points, exhibiting a higher relative error of 25\%, which can lead to increased variability and errors at mid to high charge levels. In contrast, the averaging method demonstrates greater stability across the entire charge range. With its reduced sensitivity to saturation and a lower relative error of 15\% at the saturation point, the averaging method may be the preferred and safest choice for reconstructing charges in the FASER experiment.
\section{Conclusions}
In conclusion, the production ASIC for the FASER experiment's preshower detector demonstrates significant improvements in performance. Two calibration methods—fitting and averaging—were evaluated. The new preshower detector, equipped with its modules and silicon monolithic ASICs, is scheduled for installation in FASER at the end of 2024, marking a key milestone in enhancing the experiment's detection capabilities.
\acknowledgments
The development and construction of the W-Si preshower of the FASER experiment was funded by the Swiss National Science Foundation (SNSF) under the FLARE grant 20FL21 - 201474 at the University of Geneva. Additional financial contributions from KEK, Kyushu University, Mainz University, Tsinghua University and the Heising-Simons Foundation are also acknowledged.


\end{document}